\documentclass{elsart}

\usepackage{graphicx,epsfig,epsf,amssymb}
\usepackage{amssymb}

\newcommand{\grs}{$\gamma$-rays}

\newcommand{\dg}{\ensuremath{^\circ}}
\newcommand{\hess}{H.E.S.S.}

\begin{document}

\begin{frontmatter}

\title{The Trigger System of the \hess\ Telescope Array}
  
\author[HD]{S. Funk\corauthref{add1}}
\author[HD]{, G. Hermann}
\author[HD]{, J. Hinton}
\author[HD]{, D. Berge}
\author[]{,}
\author[HD]{K. Bernl\"ohr}
\author[HD]{, W. Hofmann}
\author[LPNHE]{, P. Nayman}
\author[LPNHE]{, F. Toussenel}
\author[]{,}
\author[LPNHE]{P. Vincent}

\address[HD]{Max-Planck-Institut f\"ur Kernphysik, P.O.~Box~103980, D-69029~Heidelberg, Germany}
\address[LPNHE]{Laboratoire de Physique Nucl\'eaire et de Hautes Energies, 
  IN2P3/CNRS, Universit\'es Paris VI~\&~VII,
  4~place~Jussieu, F-75252~Paris~Cedex~05, France}

\corauth[add1]{Corresponding author, Stefan.Funk@mpi-hd.mpg.de}

\begin{abstract}

  \hess\ -- The High Energy Stereoscopic System-- is a new system of
  large atmospheric Cherenkov telescopes for GeV/TeV $\gamma$-ray astronomy. 
  This paper describes the trigger system of \hess\ with emphasis on the
  multi-telescope array level trigger. The system trigger requires the
  simultaneous detection of air-showers by several telescopes at the
  hardware level. This requirement allows a suppression of background 
  events which in turn leads to a lower system energy threshold for
  the detection of \grs. The implementation of the \hess\ trigger system
  is presented along with data taken to characterise its performance.
  
\end{abstract}


\end{frontmatter}

\section{Introduction}
\label{intro}

Present instruments in the field of ground-based $\gamma$-ray
astronomy are sensitive to photons with energies above
$\sim$100~GeV. The most sensitive of these instruments are Imaging
Atmospheric Cherenkov Telescopes (IACTs) or arrays of such
telescopes. IACTs image the Cherenkov light emitted by atmospheric
particle showers initiated by cosmic $\gamma$-rays or hadrons.  Fast
photon detectors and electronics are required to distinguish the
Cherenkov light from such air-showers from fluctuations of the night sky background
light (NSB).  Photomultiplier tubes (PMTs) are currently the most
appropriate light sensors for IACT cameras.  For $\gamma$-ray
initiated showers the light yield at the ground is roughly
proportional to the energy of the primary particle.
$\gamma$-ray showers are observed against a background of much more
numerous hadronic showers. This background can be greatly reduced
using the morphology of air-shower images. Image analysis provides 
estimates of the direction and energy of the primary
particle.

The pioneering experiment in this field was the Whipple
telescope~\cite{WHIPPLE}, which achieved an energy threshold of around 350~GeV. 
Significant improvement in sensitivity and 
energy resolution of the IACT technique was achieved in the last
generation of instruments by HEGRA~\cite{HEGRA} through the
introduction of stereoscopy, where showers are imaged simultaneously by
multiple Cherenkov telescopes. This technique provides a more accurate
measurement of shower parameters such as energy and direction and leads to an
improved $\gamma$-ray sensitivity. The current
generation of instruments aims to lower the energy threshold to below 
100~GeV. All new major installations are based on
the stereoscopic approach (\hess\ \cite{HESSProject}, VERITAS~\cite{VERITAS} and
CANGAROO-III~\cite{CANGAROO3}) or plan to adopt stereoscopy (MAGIC, phase
II \cite{MAGIC}). 

The High Energy Stereoscopic System (\hess) is located in Namibia
(Africa) at an altitude of 1800 metres. Phase~I consists of four
Cherenkov telescopes, each of 107~m$^{2}$ mirror area \cite{HESSOPT}. 
The telescopes are arranged on a square with 120 m~sides. The cameras of the
telescopes are each equipped with 960 pixels, covering a 5\dg\ field of
view with a pixel size of 0.16\dg\ \cite{HESSCAMERA}. 
The telescopes of the system were commissioned between June 2002 and December 2003.
The trigger system of \hess\ is designed to make optimum use of the
stereoscopic approach. Simultaneous observation of an air-shower 
in multiple telescopes is required at the hardware level. This
coincidence requirement reduces the rate of background events 
(for example single muons) significantly, enabling a reduction
of the energy threshold of the system by a factor $\sim$2 
compared to single telescope operation.

This paper first discusses the triggering of Cherenkov telescopes in general, 
followed by a description of the \hess\ trigger system. Technical
measurements using air-showers are presented along with a discussion
of the impact of the stereo trigger on the physics performance of the \hess\ experiment.

\section{Trigger and Readout of Cherenkov Telescopes}
\label{Sec::Triggering}

\subsection{Triggering Cherenkov Telescopes}

The triggering of Cherenkov telescopes makes use of the 
extremely short duration (a few nanoseconds) of the
Cherenkov light signal from air-showers. A typical 
requirement for triggering the readout of a telescope
is that a minimum number of pixels exhibit a signal
larger than a given threshold (typically a few photoelectrons) 
within a short time window, to reduce random triggers
from the night sky background. Following a trigger, signals are digitised
and read out, resulting in a dead-time ranging from a few 10 $\mu$s to
a few 10~ms, depending on the design of the data acquisition system.

The trigger rate (and therefore the dead-time) of a single 
telescope system is dominated by background events. 
For instruments with energy thresholds in the TeV range, hadronic 
air-showers produce the majority of this background. At the lower 
energy thresholds reached by telescopes of larger mirror area, 
single muons passing close to the Cherenkov telescope become a 
sizable part of the background.

In a system of telescopes, the requirement that two telescopes
(separated e.g. by $\sim$100~m) both trigger within a short time window
leads to a significant reduction in the rate of background events.
Since hadronic showers have a more inhomogeneous light pool, the coincidence
requirement disfavours such events in comparison to $\gamma$-ray
events. Single muons are almost completely rejected by a telescope 
multiplicity requirement.

Given this reduction in background and the advantages of 
stereoscopic reconstruction (see for example~\cite{HEGRA501}), 
it is necessary to select multiple-telescope events at least
for off-line analysis. In a system with non-negligible read-out dead-time
it is desirable to select coincident events at the hardware level.
A system level coincidence trigger has the additional advantages of 
greatly reducing the network, disc space, and CPU time requirements 
of the system. For these reasons a system level trigger was 
incorporated in the design of the \hess\ project.

During commissioning of phase I of the \hess\ telescope system, 
with two telescopes operational, some data were
taken with independent telescope triggers and software matching of
events based on GPS time-stamps (\emph{software stereo data}).
The system level trigger has been used for all data taking since
July 2003.

\subsection{Trigger and Readout of the \hess\ cameras}
\label{sec:camera}

The trigger of the \hess\ cameras~\cite{HESSCAMERA} is derived from a
multiplicity trigger within overlapping \emph{sectors}, each
containing 64 pixels.  A camera trigger occurs if the signals in $M$
pixels within a sector (\emph{sector threshold}) exceed a threshold of
$N$ photoelectrons (\emph{pixel threshold}).
The time-window for the multiplicity trigger is dictated by the minimum  
integrated charge over a programmable threshold. For a 
typical PMT pulse shape the effective trigger window is 1.3~ns 
(with a jitter of 0.14~ns). This rather narrow gate is possible 
due to the sorting of PMTs by high voltage within the camera.
This sorting minimises the time dispersion introduced by different PMT
transit times. The narrow gate guarantees maximum NSB suppression.

The trigger sectors overlap to ensure a homogeneous trigger efficiency
across the field-of-view of the camera. The sector and pixel
thresholds are programmable and are in the range of a few
photoelectrons for a minimum multiplicity of typically 2 to 4 pixels.
The PMT signals are sampled using 1~GHz Analogue Ring Samplers
(ARS)~\cite{ARS} with a ring buffer depth of 128 cells. Following a
camera trigger, the ring buffer is stopped and the content of the ring
buffer, within a programmable time window (normally 16~ns) around the
signal, is digitised, summed and written to an FPGA buffer. This
process takes $\approx$ 273 $\mu$s. During the first 10 $\mu$s, it is
possible to interrupt the digitisation with an external reset signal.
In this case, the readout of the analogue memories is stopped and the
sampling restarted. After digitisation, the transfer of all buffered
data into a global FiFo memory requires another 141 $\mu$s. The total
dead-time, including interrupt handling and data acquisition
preprocessing is 446~$\mu$s for an event that is read out, or
5.5~$\mu$s for an event that is discarded.

After read out, the camera is ready for the next event.  Further data
processing within the camera is done asynchronously, including the
transmission of data via optical fibre to the PC processor farm
located in the control building.

\section{The System Trigger of \hess}

\subsection{Overview}

The trigger of \hess\ is a two-level system.  At the first level,
telescopes independently form \emph{local triggers} (see
section~\ref{sec:camera}), at the second level a multi-telescope
coincidence decision is made by the central trigger system (CTS).  In
addition to the formation of the system level trigger, the CTS is
responsible for dead-time measurement and event synchronisation.

The \hess\ CTS consists of hardware in a central station located in
the control building of the array and of interface modules located in
each camera. The central station is designed to serve up to 8
telescopes arranged into arbitrary sub-arrays, to accomodate expansion
of the H.E.S.S. telescope system beyond phase I. The communication
between the central station and the interface modules is done via an
optical fibre system built from standard components with one duplex
pair per telescope.

\begin{figure}
\begin{center}
\mbox{\epsfig{file=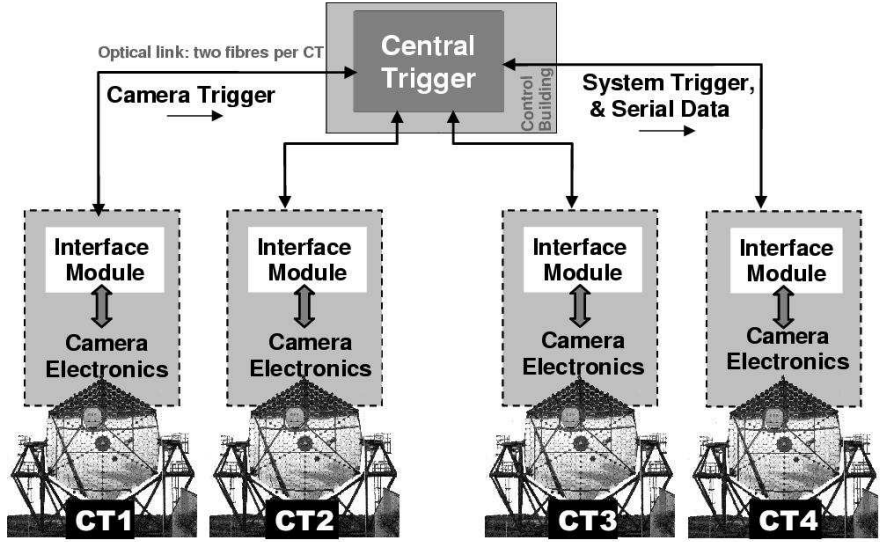, width=0.86\textwidth }}
\end{center}
\caption{Schematic of the data flow in of the \hess\ central trigger system.}
\label{fig:DataFlow}
\end{figure}

Information on all telescope triggers arrives at the central station.
If a valid coincidence occurs, the central station distributes this
information to all telescopes and the cameras of those telescopes
which participated in this system event are read out. To enable the
measurement of the system dead-time, cameras provide their current
readout status together with the trigger signal (i.e. whether the
camera is \emph{busy} acquiring a previous event or \emph{active} and
able to read out this event).  The camera trigger and readout status
information is stored for each telescope on an event by event basis in
a FiFo. From these data the dead-time of the system can be
derived. The data flow between the cameras and the central trigger
system is shown schematically in figure \ref{fig:DataFlow}.

To provide a synchronisation mechanism, the CTS assigns for each event
a unique, system-wide event number, which is distributed via the
trigger hardware to all telescopes. The event number is read by the camera 
together with the pixel data of each telescope and is used in the process of
building a system event from the individual telescope data.  An
absolute time-stamp for the system event is provided by a GPS clock in
the central station. 

\subsection{Camera Interface}

\begin{figure}
\begin{center}
\mbox{\epsfig{file=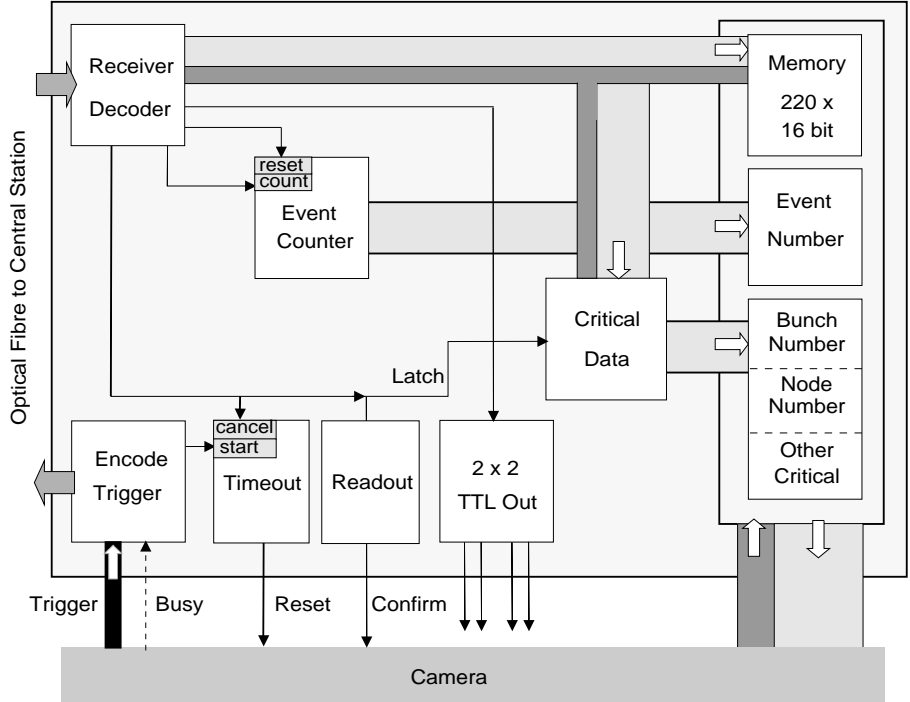, width=0.86\textwidth }}
\end{center}
\caption{Schematic illustration of the camera interface module 
of the \hess\ system trigger. The trigger path is shown in black,
data flow is shown in pale grey and 
addressing in darker grey. Pulses are shown as solid lines.}
\label{fig:LMBlock}
\end{figure}

An interface module (illustrated schematically in
figure~\ref{fig:LMBlock})
is contained within the body of each camera and is
responsible for encoding and transmission of camera triggers and
decoding and relaying pulses from the central station.  TTL trigger
pulses are received from the camera together with a TTL level
indicating the current readout status.  This information is
transformed into a width-encoded pulse which is sent via
optical fibre to the central station. 

Information from the central station is received via a second fibre.
Two transmission types are used: a system trigger pulse and a 
serial data word used to control the interface module and pass
data to the camera. 
The system trigger is also pulse width encoded: 
the telescopes that have sent active triggers receive \emph{readout} pulses, 
and all other telescopes receive a \emph{count} pulse. 
In either case a local event counter is incremented in the 
interface module and is, therefore, synchronised to the system event counter.
In the case of a \emph{readout} pulse, the event counter is 
latched into an output register on the interface module, where it is read
by the camera electronics and written into the data stream. 

If a telescope sends an active trigger but no system coincidence occurs, 
the interface module sends a reset signal to the camera after an adjustable 
delay, dictated by
the round-trip time of pulses between the telescopes and the control
building (4.2~$\mu$s) plus the time required to make a coincidence
decision (330~ns).  If a reset signal is sent, the camera discards the
event and is immediately ready for a new trigger and readout.

The serial transmission mechanism is used to periodically reset the
event counter and update synchronously a \emph{bunch number} and a 
\emph{node number} in the interface modules of all telescopes. 
Both numbers are latched into the memory of the interface module by 
the arrival of a readout pulse.
The node number identifies to which node of the computer farm data should 
be sent for processing.
The combination of the event counter and bunch number provides
a unique system wide event identifier which is used for event-building
at the farm computer specified as the receiving node.

In case a count pulse is lost, the bunch number and the reset
of the event counter ensure that the synchronisation of events is
recovered after a few seconds.

\subsection{Central Station}

The central station hardware consists of the following custom-built 
VME modules:

\begin{itemize}
\item Optical/TTL converter module
\item Programmable delay module
\item Coincidence trigger module
\item Telescope trigger scaler 
\end{itemize}

together with a GPS clock for absolute timing. 

Incoming pulses from the optical fibre are converted in the optical
converter to TTL pulses. The programmable delay 
(up to 1 $\mu$s in steps of 1~ns) compensates for the 
(fixed) delays due to different optical fibre lengths and the (varying) 
differences in the arrival times of the Cherenkov light front 
at the individual telescopes. Calculation of the variable component
is based on the current pointing direction of the telescope system.

\begin{figure}
\begin{center}
\mbox{\epsfig{file=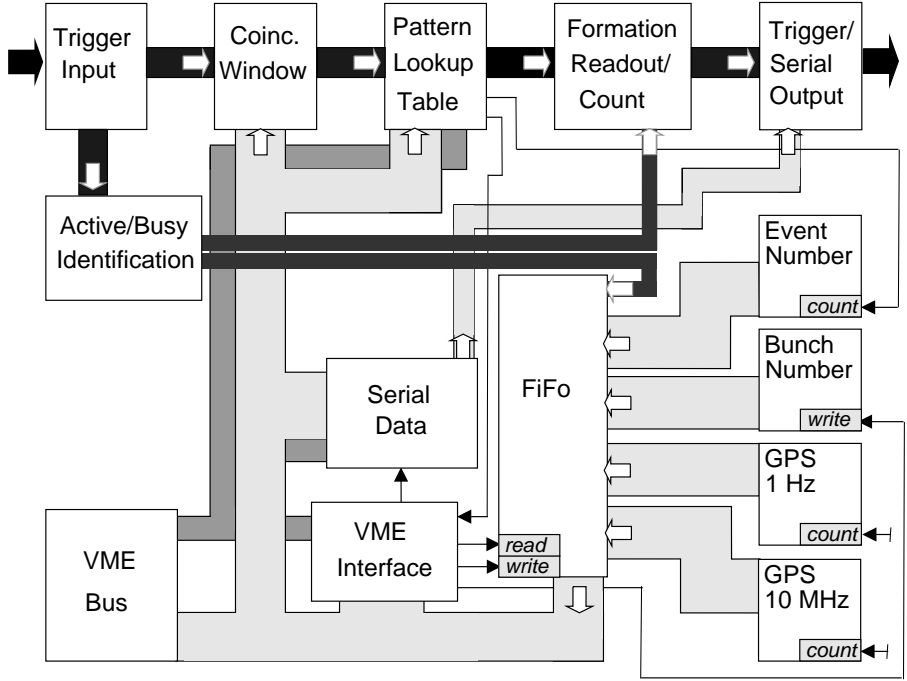, width=0.86\textwidth }}
\end{center}
\caption{The flow of information in the \hess\ central coincidence module.
The two trigger paths are shown in black. Data flow is shown in pale grey and
addressing in darker grey. Pulses are shown as solid lines.}
\label{fig:CTrigBlock}
\end{figure}

After this delay compensation, telescope triggers enter the
coincidence module where they are duplicated and follow two paths (see
figure \ref{fig:CTrigBlock}).  The coincidence path checks if the
system trigger condition is fulfilled within the 
coincidence window, without regard to the busy
status of the telescope. The coincidence window is generated by
adjusting the width of the incoming pulses to a programmable
duration and requiring a minimum overlap of 10 ns. 
The trigger condition is checked using a programmable lookup table in
which all allowed coincidence patterns are stored.  This system has
the advantage that independent sub-arrays of telescopes can be served
simultaneously. The second path identifies the active or busy
status of the telescopes and writes this information (together with
the event and bunch numbers and a GPS time-stamp) into a FiFo for
every system coincidence. Once this information has been written to
the FiFo, the trigger module is ready to accept new
triggers, about 330~ns after the coincidence occurred. 
The FiFo has a depth of 16000 events and is read out
asynchronously.  The readout time per event is 4.5~$\mu$s, 
allowing for a maximum sustained rate of 200~kHz. 
When a system trigger occurs, \emph{readout} and \emph{count} pulses
are sent via the optical converter module to the interface modules in
each camera.

Event time-stamps are provided by a commercial \emph{GPS167 Meinberg} 
GPS receiver. This clock provides  1~Hz and 10~MHz TTL signals synchronised to 
the UTC second. These signals are used as inputs for the corresponding 
counters in the coincidence module to provided relative timing.
An absolute reference time is obtained by serial readout of a complete
date/time string from the clock via RS232. The precision of the system 
has been verified by simultaneous operation of two such clocks and
using the optical pulsed emission of the Crab Pulsar as an absolute
reference~\cite{FRANZEN}. The long term accuracy of the system is $<2\mu$s.

Only coincident events are recorded by the coincidence module.
Individual telescope trigger rates (both \emph{active} and \emph{busy}) 
are monitored separately using a custom-built VME scaler.

Extensive laboratory tests of the central trigger 
system showed error rates of less than $10^{-8}$ in all aspects of system 
operation, including serial data transmission over optical fibre and 
trigger pattern identification and readout.

\section{System Trigger Characterisation}

Since December 2003, the \hess\ detector has been operating as a 
four telescope system using the system trigger described 
here\footnote{The system level trigger was installed in July 2003 
and operated until December 2003 with 2-3 telescopes, and
since then with the complete 4-telescope system.}.
During March 2004, a number of technical measurements were
made with the aim of characterising the complete \hess\ trigger.
These measurements used hadronic air-shower events to provide an
end-to-end test of the system and are described in detail below.

\subsection{Telescope Delays and Coincidence Window}

An efficient and unbiased multi-telescope trigger must provide a
coincidence window wide enough that no valid Cherenkov coincidences 
are lost. On the other hand, the window should be narrow enough to avoid
an unacceptable rate of random telescope coincidences.
The minimum achievable window is dictated by the intrinsic spread in
the arrival times of telescope triggers at the central station. 
This spread results mainly from the width and curvature of the 
Cherenkov wave front and the field-of-view of the cameras, 
and for \hess\ has an r.m.s.\ of close to 10~ns. 
The telescope trigger delay compensation -- which depends on the
pointing direction -- is updated frequently enough (in steps of 1~ns)
that there is no additional contribution to this spread.

\begin{figure}[h]
\begin{center}
\mbox{\epsfig{file=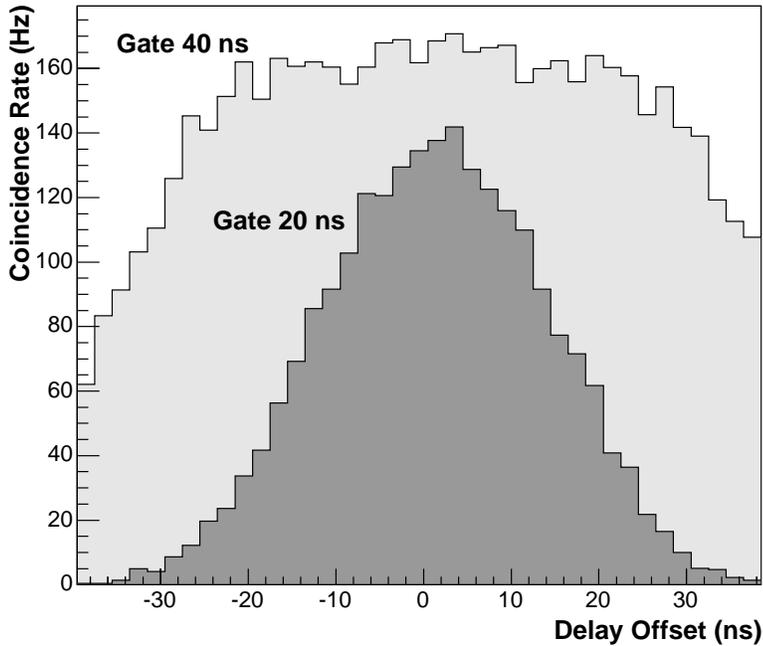, width=0.86\textwidth }}
\end{center}
\caption{System rate versus relative delay between a pair of telescopes
for coincidence windows of 20~ns and 40~ns.} 
\label{fig:DelaySweep}
\end{figure}

To demonstrate the correct calculation of the compensation delays, the
following procedure was applied to all pairs of telescopes.
Air-shower data was taken with telescopes tracking an 
astronomical target. A rather narrow coincidence window was
used and a varying offset was artificially added to the programmed
delays. The resulting curve of system rate
versus delay offset for one pair of telescopes is
shown in figure~\ref{fig:DelaySweep} for two different coincidence
windows.
 
For the 20~ns window, the delay offset which results in a maximum system rate
is $1.5\pm0.5$~ns. The 40~ns window is significantly 
wider than the intrinsic spread in arrival times ($\approx$ 10~ns rms)
and hence a clear plateau is evident, where the system rate is 
insensitive to an offset in the telescope delays. 

To avoid any bias from the system trigger on the selection of showers,
the operating coincidence gate is set to 80~ns. For typical individual
telescope rates of 1~kHz, this window introduces an acceptable 
accidental coincidence rate of 1~Hz for the 4-telescope system.

\subsection{Dead-time Determination}

For the determination of spectra and fluxes of astrophysical
$\gamma$-ray sources, an accurate measurement of the system dead-time is
required.  Every Cherenkov event triggering the \hess\ array is
recorded by the central trigger system - regardless of the readout
status of the cameras.  The CTS stores information on which telescopes
were able to provide data for a given event, as well as on those that 
triggered but were busy with the readout of a previous event. 
This information is used to determine the system dead-time.

\begin{figure}
\begin{center}
\mbox{\epsfig{file=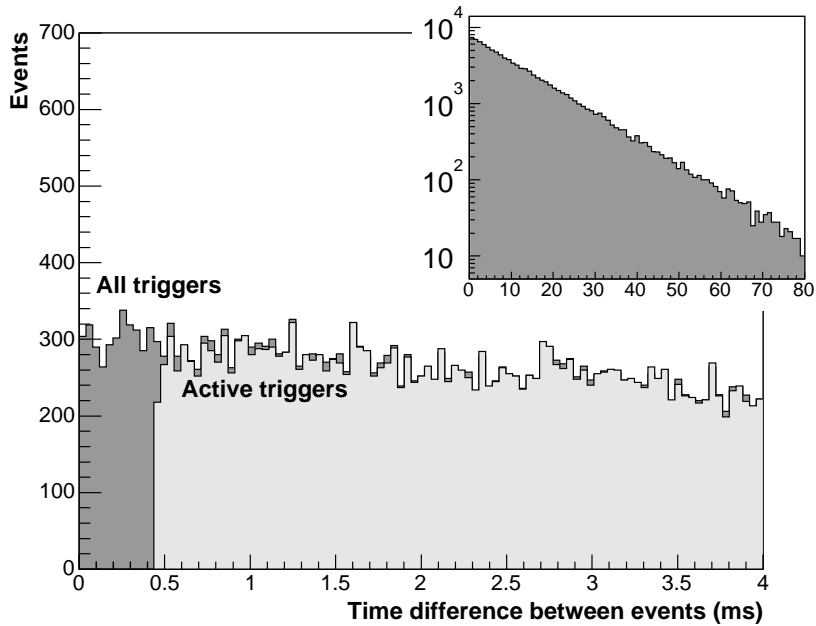, width=0.86\textwidth }}
\end{center}
\caption{
  Distribution of time differences between consecutive
  events for a single telescope. The dark grey histogram shows all 
  events triggering the system, the light grey histogram shows only 
  events for which the telescope 
  was read out. The inset shows the same distributions but with a 
  logarithmic scale to highlight the exponential behaviour.
}
\label{fig:MuonRun}
\end{figure}
 
Figure \ref{fig:MuonRun} shows the distribution of time differences
between consecutive events in a run involving only a single telescope. 
The distribution of all triggers is well described by an
exponential as expected. The distribution of events where the
telescope was read out is also exponential but with a sharp
cut-off at the camera readout dead-time of 446~$\mu$s. 

\subsection{Trigger Threshold and Trigger Rates}
\label{sec:threshold}

The selection of the trigger threshold has direct
implications for the energy threshold of the array. A low energy
threshold is clearly desirable for astrophysical reasons, however the
trigger threshold must be set high enough that the trigger rate is
both stable and manageable. In the low threshold regime where night sky
background triggers dominate, the rate is expected to be both very
high and unstable. Measurements of the dependence of the system trigger
rate on all available parameters dictating the threshold are therefore
crucial in determining the optimal operating parameters.  

The adjustable parameters which directly affect the 
energy threshold and trigger rate of the system are: the camera
pixel and sector thresholds (see section~\ref{sec:camera}), and 
the telescope multiplicity requirement. The measurements
described below involve variation of these parameters. All
were made with all four telescopes of the system tracking a relatively
dark region of the sky close to zenith.

\begin{figure}
\begin{center}
\mbox{\epsfig{file=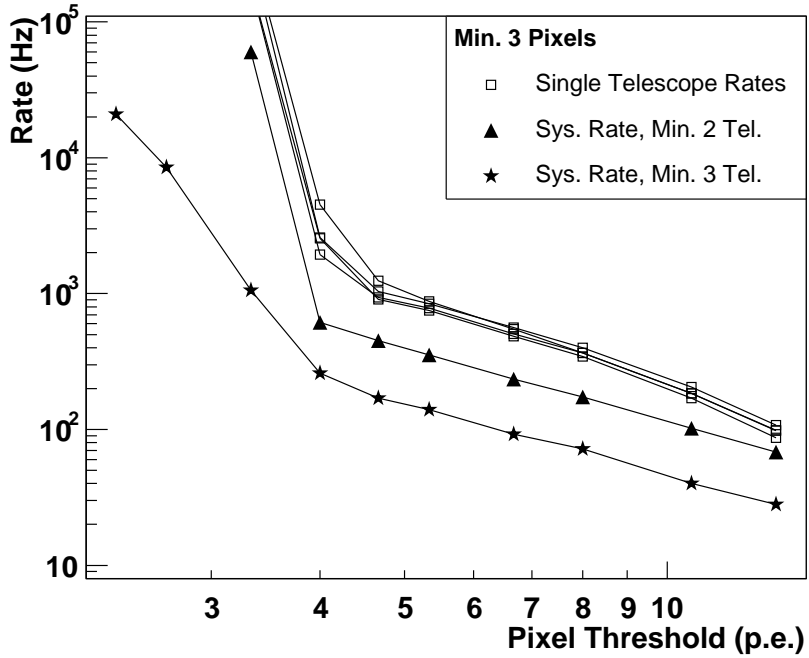, width=0.86\textwidth}}
\end{center}
\caption{Single telescope rates and system rate (for a sector threshold
  of 3 pixels) against pixel threshold. The system rate is
  shown for telescope multiplicities 2 and 3. Statistical
  error bars are in all cases smaller than the symbols.}
\label{fig:RateVsPixel}
\end{figure}

Figure~\ref{fig:RateVsPixel} shows how the
rates of individual telescopes and the system rate depend on the pixel
threshold. Curves are shown for minimum telescope multiplicities of 2
and 3. The pixel threshold is quoted in units of photoelectrons (p.e.),
defined as the mean amplitude of a single photoelectron
signal at the pixel comparator (21~mV).
In all curves, two regimes are clearly present: a smooth power law
dependence at higher thresholds and a rapid increase in rate below
$\approx 4$~p.e.  For low pixel thresholds, the camera triggers are
dominated by pixel coincidences due to night sky background
fluctuations and the system rate is dominated by accidental telescope
coincidences.  For higher pixel thresholds, random pixel coincidences
in the cameras are rare and Cherenkov events from cosmic ray
air-showers dominate.

The small dispersion in single telescope rates (in the air-shower
dominated regime) that is evident in figure~\ref{fig:RateVsPixel}
demonstrates the homogeneity of the array.  Inter-telescope
differences in mirror reflectivity and pixel efficiency are at a level
of less than 10\%.

\begin{figure}
\begin{center}
  \mbox{\epsfig{file=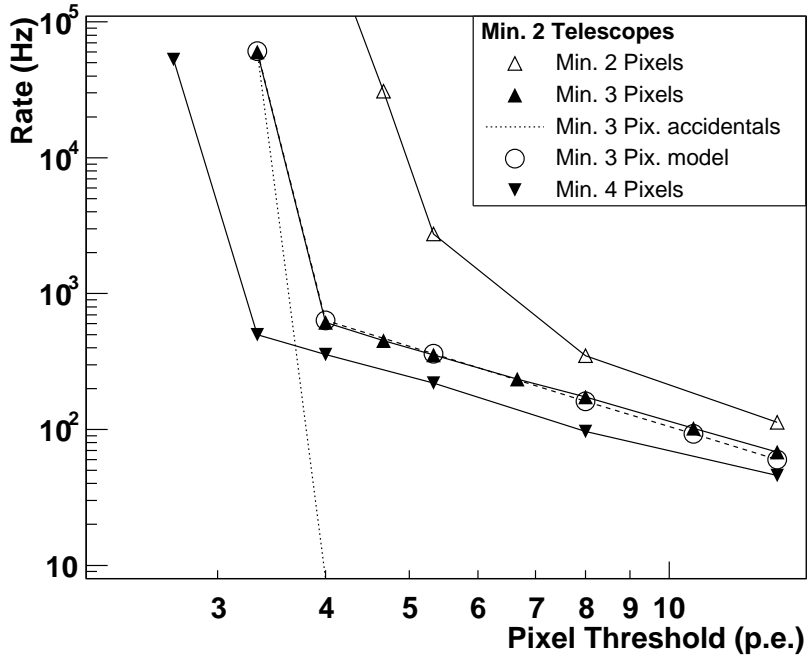, width=0.86\textwidth}}
\end{center}
\caption{System trigger rate against pixel threshold for three sector
  thresholds.  For a sector threshold of 3 pixels the
  predicted rate from accidental telescope coincidences is
  shown along with the simulated rate for this configuration
  (open circles). The statistical error bars are smaller than the symbols
  for all measured curves.}
\label{fig:RateVsPixel2}
\end{figure}

Figure \ref{fig:RateVsPixel2} shows a comparison of
system rate versus pixel threshold for three sector thresholds
(pixel multiplicities).
For a threshold of 2 pixels, the transition between the NSB 
and air-shower regimes is rather gradual. For such a condition the
system trigger may be affected by NSB fluctuations for a wide range of
possible pixel thresholds. As expected, at a higher sector threshold
of 4 pixels, a lower operating pixel threshold is attainable. 
For sector multiplicities of 3 and 4,
similar minimum image sizes and system trigger rates are attainable
(within the air-shower dominated regime). As a sector threshold of 
3 preferentially selects the more compact images typical of gamma-ray
initiated showers, this threshold is preferred.

Figure \ref{fig:RateVsPixel2} also shows the expected rate of
accidental coincidences for a sector threshold of 3 pixels,
calculated using the measured single telescope rates and the
system coincidence window of 80~ns. The measured system trigger rate at a
3.4~photoelectron threshold is in good agreement with the
predicted rate of accidentals, indicating that the system 
trigger behaves as expected, even at this extremely high rate
(50~kHz). The open circles show the system trigger rate predicted
by Monte Carlo simulations using the CORSIKA~\cite{CORSIKA} air-shower
simulation and an optical ray tracing and electronics simulation of
the \hess\ telescopes~\cite{SIMHESSARRAY} (combined with the expected
accidental rate). The comparison of the predicted and measured rates 
provides an end-to-end test of the simulations and the good agreement
is encouraging.

The measurements described above were all made in a relatively 
dark region of the sky, away from the galactic plane. In such regions
the median pixel NSB determined from pixel
currents~\cite{HESS_Calib} is close to
$9.2\times\,10^{7}$ photoelectrons s$^{-1}$ per pixel, compared with the
expected value of $8.5\pm1.3\,\times\,10^{7}$ photoelectrons s$^{-1}$ per 
pixel~\cite{HESS_NSB}. 
However, as $\gamma$-ray 
observations of all parts of the sky are planned with \hess, 
the system must operate stably in much brighter regions of the sky. 
In a field-of-view with many bright stars or with increased diffuse
emission, NSB triggers may occur at higher
pixel threshold than in a dark region. To study this effect, the rate 
versus threshold measurements described above were repeated in the 
particularly bright region around $\eta$ Carinae. In this region the
transition to NSB dominance occurs close to 4.7~p.e. for a sector 
threshold of 3 pixels. 

To avoid significant NSB trigger rates in bright regions of the sky
and as a compromise between higher $\gamma$-ray collection efficiency
and reduced energy threshold, a configuration of sector threshold 3,
pixel threshold 5.3, and telescope multiplicity 2 was chosen for
\hess\ observations.

\subsection{Zenith Angle Dependence}

The measurements described in the previous section were performed close to
zenith. Standard \hess\ observations take place in a zenith angle range of
0-35\dg. In certain circumstances this range is extended as far as
60\dg. At large zenith angles, air-showers are observed through a much
greater atmospheric column depth. Cherenkov photons from such showers
suffer more from scattering and absorption and have a larger (and
dimmer) footprint on the ground. As a consequence of the reduced
density of photons in the light pool, the effective energy threshold of
the system is increased. On the other hand, the larger light pool
diameter results in an improved effective collection area.  

\begin{figure}
\begin{center}
\mbox{\epsfig{file=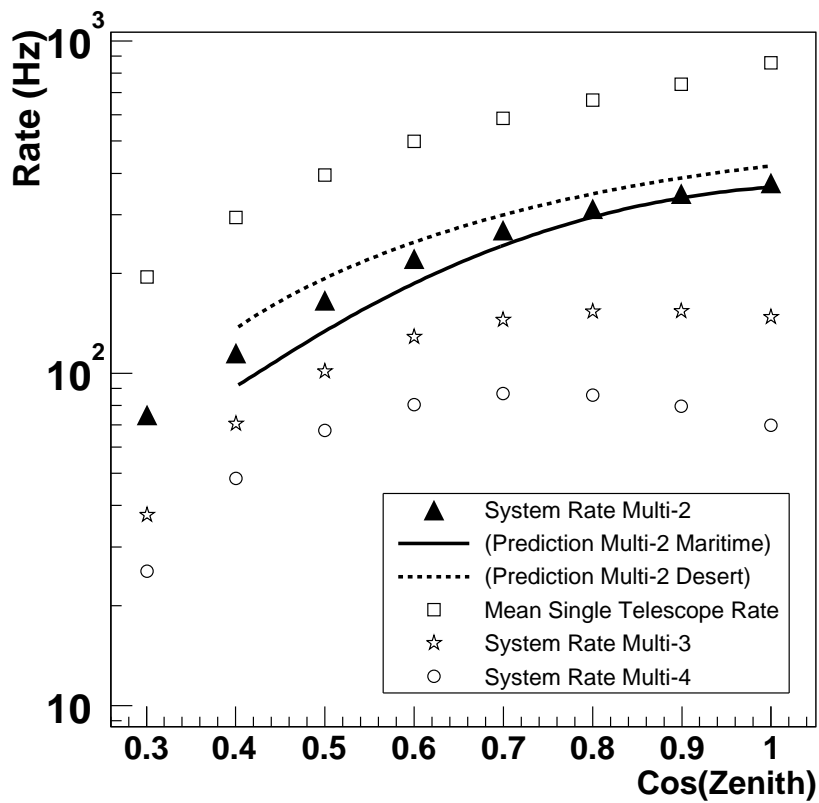,width=0.86\textwidth }}
\end{center}
\caption{Mean single telescope and system rate against cosine zenith
  for telescope multiplicities of 2, 3 and 4. The solid and dashed lines show the rate predicted 
  for multiplicity-2 using two different atmospheric models. 
  These data were taken with telescopes pointing to the north. 
 All statistical error bars are smaller than the plotted symbols.}
\label{fig:RateVsZenith}
\end{figure}

Figure~\ref{fig:RateVsZenith} shows the rate of the system (and the
mean single telescope rate) as a function of the cosine of the zenith
angle. A smooth, monotonic decrease is evident for the single telescope and
multiplicity-2 rates as expected from the increasing absorption of
showers and the steep energy  
spectrum of the hadronic background.  The solid and dashed lines
in figure~\ref{fig:RateVsZenith} shows the predicted behaviour
assuming two different atmospheric transmission tables 
calculated using MODTRAN~\cite{ATMOSPHERE}.
The solid line is based on a rather conservative assumption of aerosol content 
(maritime haze, boundary layer starting at sea level). The dotted line corresponds to 
a clearer atmosphere (desert haze, boundary layer starting at 1800~m).
The difference between the two predictions indicates the current uncertainty in our
understanding of the atmosphere at the \hess\ site. 

The increased diameter of the Cherenkov light pool at large zenith
angles leads to an increased average telescope multiplicity. This
effect is apparent from a comparison of the telescope multiplicity-2
rate with the curves for larger telescope multiplicity 
in figure~\ref{fig:RateVsZenith}, indeed the multiplicity-4 rate 
reaches its maximum at 40\dg\ zenith angle.

\subsection{Convergent Telescope Pointing}

\begin{figure}
\begin{center}
\mbox{\epsfig{file=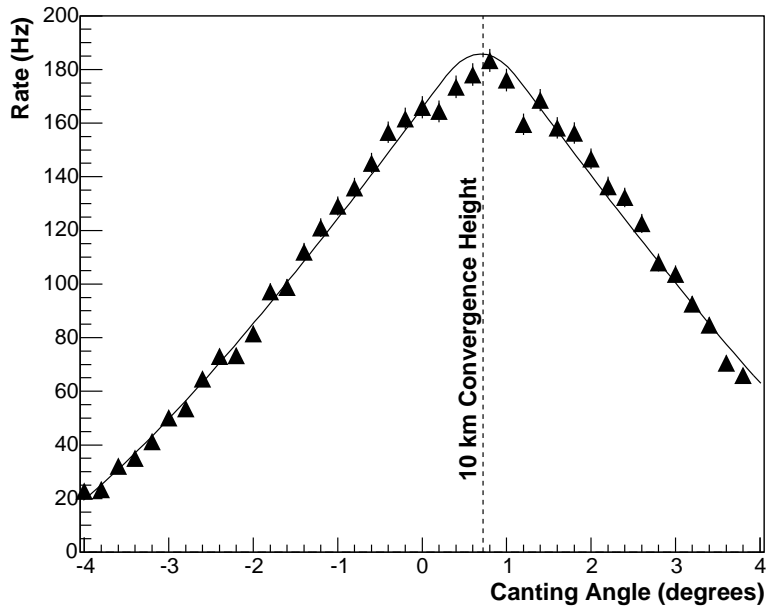,width=0.86\textwidth }}
\end{center}
\caption{System rate as a function of the relative alignment of the 
  telescope axes for two telescopes separated by 120~m.
  The inclination of the telescopes takes place in the plane
  connecting the two telescopes. Positive values refer to
  convergent telescope pointing. The solid line is a fit to the 
  data of a simple geometrical model.}
\label{fig:Canting}
\end{figure}

The simplest pointing strategy for a telescope array is to track a
target with all telescope axes parallel. All measurements described
above employed this observation mode. However, given a typical
emission height of Cherenkov photons $\sim$10~km above the observation
level, a more effective strategy may be to \emph{cant} the telescopes
towards each other slightly \cite{HEGRACANT}. Given the altitude and array spacing of
\hess, the convergent angle required to maximise the overlap of the
telescope field-of-views at the height of maximum shower development
is around 0.7\dg. Figure~\ref{fig:Canting} shows the measured
dependence of system trigger rate on the canting angle between
telescopes.  The maximum trigger rate occurs at a canting angle of
0.65\dg, corresponding to convergence at a point 10.5~km above the
\hess\ site. The solid curve shows a fit of a simple geometrical model
of overlapping telescope field-of-views.

Maximising the overlap of telescope field-of-views results in
$\gamma$-ray images that typically lie closer to the centre of 
the cameras and increases the average number of usable images
in the analysis. This in turn leads to improved angular
resolution and hence to better sensitivity. 
Convergence to a fixed atmospheric depth of 270~g$\,$cm$^{-2}$ 
(the depth of maximum Cherenkov photon emission for an
average 100~GeV $\gamma$-ray shower) is planned for future 
\hess\ observations.

\section{Implications for System Performance}

\begin{figure}[t]
\begin{center}
\mbox{\epsfig{file=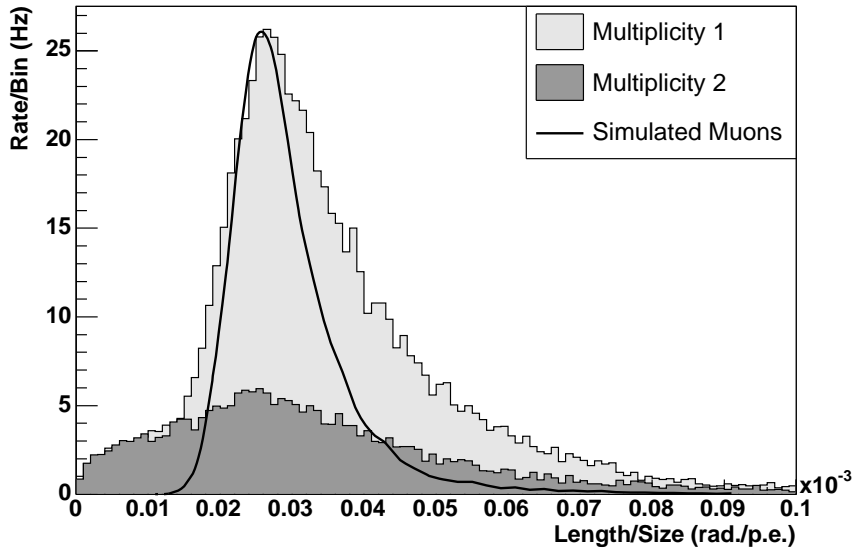, width=0.86\textwidth }}
\end{center}
\caption{Distribution of \emph{length}/\emph{size} of images 
  for telescope multiplicity 1 and 2. The distribution of 
  simulated single muons is shown by the solid curve.}
\label{fig:LoverS}
\end{figure}

The operation of a multiplicity-2 telescope trigger for the \hess\
system has many implications. A major consequence is the removal of
single muon events at the trigger level. Single muon images have a
characteristically narrow distribution of surface brightness. As a
consequence they are limited to a small region in the parameter (image
\emph{length})/(image \emph{size}) (following the definition of image parameters of
Hillas~\cite{HILLAS}). Figure~\ref{fig:LoverS} shows the distribution
of $length/size$ with and without a telescope multiplicity requirement
for fixed camera trigger conditions. For telescope multiplicity-1
there is a large peak at around $3\times10^{-5}$
radians/photoelectron. This peak is produced by images of single
muons as can be seen by comparison with the curve for simulated muons
shown in figure~\ref{fig:LoverS}. The tail to the right of the peak is
likely produced by more complex shower images that are still dominated
by a single muon.
From the telescope multiplicity-2 curve in figure~\ref{fig:LoverS}
it is clear that a multi-telescope trigger is extremely effective
at removing such background events; the peak attributed to single
muon images is almost completely absent.
         
\begin{figure}
\begin{center}
\mbox{\epsfig{file=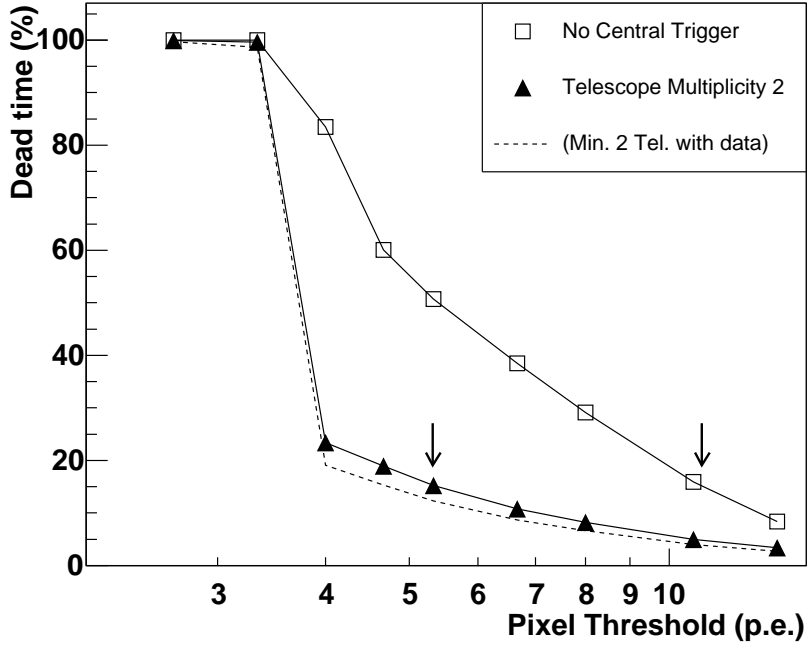, width=0.86\textwidth }}
\end{center}
\caption{
  Derived system dead-time as a function of pixel trigger threshold (for
  a fixed sector threshold of 3 pixels), with 
  and without a system level trigger. The dashed and solid curves are calculated 
  using different definitions of the system dead-time (see text).
  The left arrow shows the current operating pixel threshold of \hess,
  the right arrow shows the threshold corresponding to the same 
  system dead-time with no telescope multiplicity requirement.
} 
\label{fig:DeadThresh}
\end{figure}

The reduction in background evident in figure~\ref{fig:LoverS} leads
to a lower system trigger rate at a given threshold and hence to 
reduced readout dead-time. Figure~\ref{fig:DeadThresh} compares the
incurred system dead-time versus pixel threshold with and without a 
telescope multiplicity requirement. 

The dashed line shows the dead-time, calculated defining a \emph{live} system
trigger as one where at least 2 telescopes were read out. For the solid curve a
more conservative requirement, that all triggered telescopes were read out,
was applied. To achieve an acceptable dead-time
fraction of 10--15\% without a system level trigger, the pixel threshold 
would have to be increased by a factor $\sim$2. As the energy threshold
of the instrument scales roughly with the pixel threshold, operation of
\hess\ with a central trigger significantly increases the sensitive
energy range of the array.

During the construction and commissioning phase of \hess, $\gamma$-ray 
observations have been made using several array configurations. The 
improvements in performance associated with the development of \hess\
from a single telescope system to a full array, are illustrated in
table~\ref{tab:Crab}. The rate of $\gamma$-rays expected from the 
Crab Nebula is used as a figure of merit in this table. 
Observations of the Crab Nebula by \hess\  have been used 
to confirm the performance predicted by air-shower and 
detector simulations. The improvement in performance
between the \emph{2-Tel. Soft.} and \emph{2-Tel. Hard.}
configurations is solely a consequence of the introduction of the
system level trigger.

\begin{table}
  \begin{center}
    \begin{tabular} {|l|c|c|ccc|}\hline
      Configuration & Camera & System & \multicolumn{3}{|c|}{Crab $\gamma$-ray Rate (min$^{-1}$)} \\ 
      & Trigger & Rate & \multicolumn{2}{c}{Predicted} & Measured \\
      & Condition &  (Hz)  & Pre-cuts & Post-cuts & Post-cuts \\\hline
      Single Telescope & 6.7 pe, 4 pix & 170  & 9.1$\pm$1.6 & 3.0$\pm$0.5 & 3.8$\pm$0.1 \\
      2-Tel. Soft. & 6.7 pe, 4 pix & 50 (280) & 5.6$\pm$1.0 & 3.6$\pm$0.7 &  \\
      2-Tel. Hard. & 5.3 pe, 3 pix & 90       & 13.5$\pm$2.4 & 6.7$\pm$1.2 &  \\
      3-Tel. Hard. & 5.3 pe, 3 pix & 180      & 20.2$\pm$3.7 & 11.8$\pm$2.1 & 10.4$\pm$0.3 \\
      4-Tel. Hard. & 5.3 pe, 3 pix & 270      & 25.9$\pm$4.7 & 15.6$\pm$2.8 &  \\\hline
    \end{tabular}
  \end{center}
  \vspace{5mm}
  \caption{
    Comparisons of the performance of different \hess\ configurations. 
    The typical system trigger rate is given together with the predicted 
    (and measured) $\gamma$-ray rate from the Crab Nebula (at 47\dg\ zenith angle).
    The predicted rates are calculated using the spectrum published by the HEGRA
    collaboration~\cite{HEGRACRAB}, the errors shown are statistical only.
    \emph{Soft.} and \emph{Hard.} refer to
    software and hardware 2-telescope multiplicity requirements. For the 
    \emph{2-Tel. Soft.} configuration the trigger rate including single telescope
    events is shown in parentheses.
  }
  \label{tab:Crab}
\end{table}

\begin{figure}
\begin{center}
\mbox{\epsfig{file=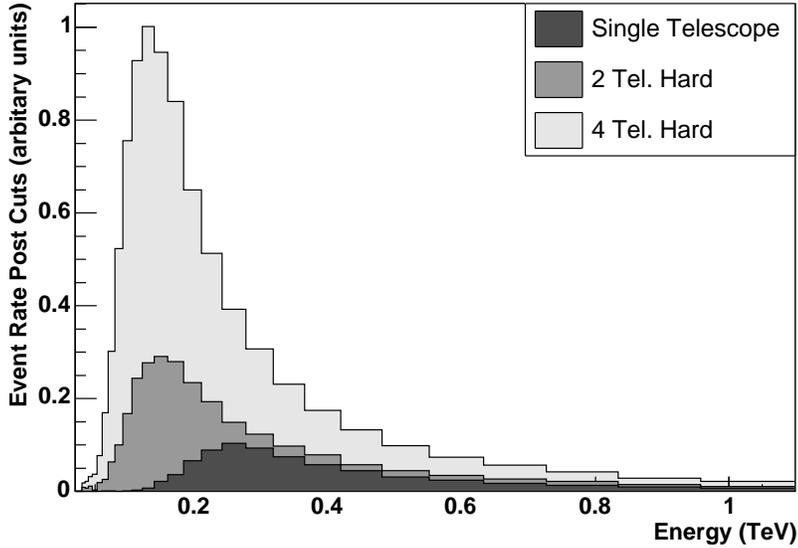,width=0.86\textwidth}}
\end{center}
\caption{
  Post-cuts differential rate expected on the basis of detailed
  simulations for a Crab-like source at zenith, 
  for different \hess\ configurations. The energy threshold is conventionally 
  defined as the peak in this distribution and is 265 GeV, 145 GeV and 125 GeV
  for the single telescope, 2-telescope and 4-telescope configurations 
  respectively.
}
\label{fig:diffrate}
\end{figure}

Figure~\ref{fig:diffrate} illustrates the decreasing energy threshold
of the array during the commissioning phase.
With the complete 4-telescope \hess\ phase-I 
system a hardware energy threshold of 100~GeV is achieved at zenith. 
After $\gamma$-ray selection cuts the threshold increases to 
125~GeV. In comparison, the post-cuts threshold
for data taken with a single \hess\ telescope is 265~GeV. For this 
configuration a \emph{length/size} cut is imposed to reject single
muons off-line, at the expense of an increased analysis threshold.

\section{Summary}

The system level trigger of the \hess\ detector provides an
effective suppression of background events at the hardware level 
and pre-selects events for stereoscopic reconstruction.
This background rejection enables the operation of
\hess\ with a lower dead-time and at a lower energy 
threshold for the detection of \grs.
The combination of large mirror area, fast camera electronics, and
stereoscopy has allowed the first phase of the \hess\ project 
to achieve its target energy threshold of 100~GeV. 

\section*{Acknowledgements}

The authors would like to acknowledge the support of their host institutions, and additionally support from
the German Ministry for Education and Research (BMBF), the French Ministry for Research, and the Astroparticle
Interdisciplinary Programme of the CNRS.
We appreciate the excellent work of the electronic engineering and technical support staff in 
Heidelberg (in particular, T.\ Schwab and N.\ Bulian), in Paris,
and in Namibia in the construction and operation of the equipment. We acknowledge
the support of S.\ Schlenker during the measurement campaign.


\begin{thebibliography}{}
  
\bibitem{WHIPPLE} Weekes, T.C. {\it et al.}, 1989, Astrophys. J. \textbf{342}, 379.
\bibitem{HEGRA} Daum, A. {\it et al.}, 1997, Astropart. Phys. \textbf{8}, 1.
\bibitem{HESSProject}
  Hinton, J.A. ({\it H.E.S.S. Collaboration}), 2004, New Astron. Rev. \textbf{48}, 331.
\bibitem{VERITAS} Weekes, T.C. {\it et al.}, 2002, Astropart. Phys. \textbf{17}, 221.
\bibitem{CANGAROO3}  Kubo, H. {\it et al.}, 2004, New Astron. Rev. \textbf{48}, 323.
\bibitem{MAGIC} Lorenz, E. ({\it MAGIC Collaboration}), 2004, New Astron. Rev. \textbf{48}, 339.
\bibitem{HESSOPT} Bernl\"ohr, K. {\it et al.}, 2003, Astropart. Phys. \textbf{20}, 111.
\bibitem{HESSCAMERA} Vincent. P. {\it et al.}, 2003, Proc. 28th ICRC (Tsukuba), Univ. Academy Press, Tokyo. p. 2887.
\bibitem{HEGRA501}  Aharonian, F.A. {\it et al.}, 1999, Astron. \& Astrophys. \textbf{349}, 11.
\bibitem{HEGRATRIGGER} Bulian, N. {\it et al.}, 1998, Astropart. Phys. \textbf{8}, 223.
\bibitem{ARS} Feinstein, F. ({\it ANTARES Collaboration}), Nucl. Instrum. Meth. A \textbf{504}, 258.
\bibitem{FRANZEN} Franzen, A. {\it et al.}, 2003, Proc. 28th ICRC (Tsukuba), Univ. Academy Press, Tokyo. p. 2987.
\bibitem{CORSIKA} Heck, D. {\it et al.}, 1998, Forschungszentrum Karlsruhe Report \textbf{FZKA 6019}.
\bibitem{SIMHESSARRAY} Bernl\"ohr, K. 2004, In preparation.
\bibitem{ATMOSPHERE} Bernl\"ohr, K. 2000, Astropart. Phys. \textbf{12}, 255.
\bibitem{HESS_Calib} Aharonian, F.A. {\it et al.}, 2004, Submitted to Astropart. Phys.
\bibitem{HESS_NSB} Preuss, S. {\it et al.}, 2002, Nucl. Instrum. Meth. A \textbf{481}, 229.
\bibitem{HEGRACANT} Lampeitl, H. {\it et al.}, 1999, AIP Proc. GeV-TeV Gamma Ray
  Astrophysics Workshop (Snowbird), \textbf{515}, 328 (astro-ph/9910461).
\bibitem{HILLAS} Hillas, A. M. 1985, Proc. 19th ICRC (La Jolla) \textbf{3}, 445.
\bibitem{HEGRACRAB} Aharonian, F.A. {\it et al.}, 2000, Astrophys. J. \textbf{539}, 317.

\end{thebibliography}
\end{document}